\documentstyle[12pt]{article}
\makeatletter
\def\fmslash{\@ifnextchar[{\fmsl@sh}{\fmsl@sh[0mu]}}
\def\fmsl@sh[#1]#2{%
  \mathchoice
    {\@fmsl@sh\displaystyle{#1}{#2}}%
    {\@fmsl@sh\textstyle{#1}{#2}}%
    {\@fmsl@sh\scriptstyle{#1}{#2}}%
    {\@fmsl@sh\scriptscriptstyle{#1}{#2}}}
\def\@fmsl@sh#1#2#3{\m@th\ooalign{$\hfil#1\mkern#2/\hfil$\crcr$#1#3$}}
\makeatother
\global\arraycolsep=2pt 
\input{epsf}
\begin{document}
\thispagestyle{empty}
\begin{titlepage}
\hskip -3cm
\begin{flushright}
{\bf TTP 95--07} \\
     March 1998  \\
     hep-ph/9803398
\end{flushright}

\vspace{1cm}

\begin{center}
{\Large\bf SPIN EFFECTS} \vspace*{5mm} \\ 
{\Large\bf IN HEAVY QUARK PROCESSES}
\end{center}

\vspace{0.8cm}

\begin{center}
{\sc Thomas Mannel}  \vspace*{2mm} \\
{\sl Institut f\"{u}r Theoretische Teilchenphysik, \\
     Universit\"at Karlsruhe \\
     D -- 76128 Karlsruhe, Germany.} 
\end{center}
\vfill
\begin{abstract}
\noindent
In the infinite mass limit for a heavy quark its spin 
decouples from the QCD dynamics, which leads to the heavy-quark 
spin symmetry. After a short discussion of spin symmetry some 
applications are considered. 
\end{abstract}
\vfill
\begin{center}
{\it Contribution to the} \\ 
{\it Crakow Epiphany Conference on Spin Effects  
     in Particle Physics}  \\
                   January 9--11, 1998, Crakow, Poland.
\end{center}
\end{titlepage}
\newpage
\section{Introduction}

Over the last ten years the heavy mass limit has become a standard 
tool in heavy quark physics \cite{reviews}. The main 
impact of the $1/m_Q$ expansion 
is that the strong interaction connecting weak processes of heavy 
quarks with those of heavy hadrons can be handled in a more 
efficient way. This is mainly due to the presence of symmetries
which appear in the heavy quark limit \cite{HQSymm} and which 
restrict the 
nonperturbative quantities severely, at least in some cases. 
Furthermore, off the symmetry limit, 
i.e.\ the limit $m_Q \to \infty$, one may compute 
or at least parametrize the corrections 
in a systematic fashion.              

These corrections are characterized by two quantities, namely 
$\alpha_s$ and $ \Lambda_{QCD} / m_Q$, both of which are small for 
a sufficiently heavy quark. The $\alpha_s$ corrections are perturbative 
and may be calculated systematically using the Feynman rules of 
Heavy Quark Effective Theory (HQET) \cite{HQET}. The other kind 
of corrections, 
the $\Lambda_{QCD} / m_Q$ corrections, are nonperturbative and are 
usually parametrized in terms of certain matrix elements. 

One of the two symmetries of the heavy mass limit is the so-called
spin symmetry, which mainly tells us that in the heavy mass limit 
the spin of the heavy quark decouples from the dynamics. The second 
symmetry of the heavy mass limit is a heavy flavour symmetry which 
allows us to replace  an infinitely heavy quark in some heavy hadron
by another infinitely heavy hadron with the same four-velocity, but 
with a different flavour, without changing anything. This is true 
because in full QCD the dependence on the quark flavour enters only 
through the different masses. 

We shall first give a brief account on the heavy flavour symmetries 
and then discuss a few applications of the heavy quark spin symmetry, 
such as the relations between 
$B \to D \ell \bar{\nu}_\ell$ and  $B \to D^* \ell \bar{\nu}_\ell$, 
the polarization of $b$ hadrons in $Z_0$ decays, some polarization 
effects in $\Lambda_c$ decays and finally the quark helicities in 
$\Lambda_b \to \Lambda \gamma$. 

\section{Heavy Quark Limit and its Symmetries}
Because of space and time limitations we shall not deduce the heavy
quark limit from QCD, rather we refer the reader to one of the numerous 
reviews which are available on this subject \cite{reviews}. 
The final result of some 
algebraic manipulations of the QCD Lagrangian and the field of the 
heavy quark is their expansion in powers of $1/m_Q$, taking a limit
in which the heavy quark velocity  $p_Q / m_Q$ is kept fixed. More 
precisely, the heavy quark momentum is split into a ``large'' part
scaling as $m_Q$ and a small residual part $k$, independent of $m_Q$, 
i.e. one writes $p_Q = m_Q v + k$. In this limit the relevant degree 
of freedom is the static heavy quark field $h_v (x)$ moving with  the 
fixed velocity $v$ and having a residual momentum 
$k h_v (x) \widehat{=} iD h_v (x)$. 
 
If $Q (x)$ is the heavy quark field of full QCD, one obtains the
$1/m_Q$ expansions 
\begin{equation} \label{FWfield}
Q(x) =  e^{-im_Q v\cdot x}\left[ 1 +\frac{1}{2m_Q}(i\fmslash{D}_{\perp}) +
\frac{1}{4m_Q^2}\left( (v\cdot D) \fmslash{D}_{\perp} - \frac{1}{2}
\fmslash{D}_{\perp}^2 \right) + \cdots \right] h_v (x) 
\end{equation}
and 
\begin{equation} \label{FWlag}
{\cal L} = \bar h_v (iv \cdot D) h_v 
              + \tilde{K}_1 + \tilde{M}_1 + \tilde{E}_1 
              + \tilde{K}_2 + \tilde{M}_2 + \tilde{E}_2 
              + \cdots 
\end{equation}
where we have defined the abbreviations
\begin{eqnarray}
\tilde{K}_1 &=& \bar{h}_v \frac{(iD)^2}{2m_Q} h_v, \quad 
\tilde{M}_1 = 
\frac{(-i)}{2m_Q}\bar{h}_v \sigma_{\mu \nu}(iD^\mu)(iD^\nu) h_v, \quad
\tilde{E}_1 = \bar{h}_v \frac{(ivD)^2}{2m_Q} h_v 
\nonumber \\
\tilde{K}_2 &=& \frac{1}{8m_Q^2}
   \bar{h}_v [(iD^\mu), [(-ivD),(iD_\mu)]] h_v \\ \nonumber 
\tilde{M}_2 &=& \frac{(-i)}{8m_Q^2}
   \bar{h}_v \sigma_{\mu \nu}\{ (iD^\mu), [ (-ivD),(iD^\nu)]\} h_v, \quad
\tilde{E}_2 = \bar{h}_v \frac{(ivD)^3}{8m_Q^2} h_v 
\end{eqnarray}
The leading term of these expansions
together with the usual Lagrangian for the light degrees 
of freedom determines the dynamics of HQET. A remarkable feature 
of the leading term of the Lagrangian is that it has two additional 
symmetries which have not been present in full QCD. 

The first symmetry which arises is a heavy flavour symmetry. 
The interaction of the 
quarks with the gluons is determined by the color quantum numbers and the 
dependence on flavour enters in full QCD only through the different 
quark masses. 
For the light quarks the fact that the light quark current masses are small 
compared to the QCD scale $\Lambda_{QCD}$ yields the well known flavour 
symmetry for the light quarks; in the heavy mass limit a flavour symmetry 
arises in a similar manner: once the heavy quark is replaced by a static 
source of colour moving with a definite velocity the flavour does not matter 
anymore. In other words, for two heavy flavours $b$ and $c$ an $SU(2)$ 
symmetry 
emerges which relates $b$ and $c$ quarks moving with the same velocity.

For the case of two heavy flavours $b$ and $c$ one has to leading order the 
Lagrangian   
\begin{equation}
{\cal L}_{heavy} = \bar{b}_v (v \cdot D) b_v + \bar{c}_v (v \cdot D) c_v ,
\end{equation}
where $b_v$ ($c_v$) is the field operator $h_v$ for the $b$ ($c$) quark 
moving with velocity $v$. 
This Lagrangian is obviously invariant under the $SU(2)_{HF}$ rotations  
\begin{equation}
\left( \begin{array}{c} b_v \\ c_v \end{array} \right) \to 
U_v \left( \begin{array}{c} b_v \\ c_v \end{array} \right) \quad 
U_v \in SU(2)_{HF} .
\end{equation}
We have put a subscript $v$ for the transformation matrix $U$, since 
this symmetry only relates heavy quarks moving with the same velocity.  

The second symmetry emerging in the heavy mass limit is the so called spin 
symmetry. To leading order both spin degrees of freedom couple in the same 
way to the gauge field. We rewrite the leading-order Lagrangian as 
\begin{equation}
{\cal L} = \bar{h}_v^{+s} (iv D) h_v^{+s} + \bar{h}_v^{-s} (iv D) h_v^{-s},
\end{equation}
where $h_v^{\pm s}$ are the projections of the heavy quark field on a 
definite spin direction $s$
\begin{equation}
h_v^{\pm s} = \frac{1}{2} (1 \pm \gamma_5 \fmslash{s}) h_v, 
\quad s\cdot v = 0 .
\end{equation}
This Lagrangian has a symmetry under the rotations of the heavy quark 
spin, which is formally again an $SU(2)_{SS}$ symmetry given by 
\begin{equation}
\left( \begin{array}{c} h_v^{+s} \\ h_v^{-s} \end{array} \right) \to 
W_v \left( \begin{array}{c}  h_v^{+s} \\ h_v^{-s} \end{array} \right) \quad 
W_v \in SU(2)_{SS} .
\end{equation}
The spin rotations may explicitely represented by 
\begin{equation}
W_v = \exp \left( -i \phi \fmslash{\epsilon} \fmslash{v} \gamma_5 \right) 
\end{equation}
where we have introduced the rotation axis $\epsilon$ satisfying 
$v\epsilon = 0$ and $\epsilon^2 = -1$ and the rotaion angle $\phi$. 
In the rest frame $v = (1,0,0,0)$ this reduces to the well known 
representation of rotations of spinors 
\begin{equation}
W_v = \exp \left( - i \phi \vec{\epsilon} \cdot \vec{\sigma} \right)
\end{equation}
where $\vec{\sigma}$ is the usual vector of the three Pauli matrices. 

Thus in the heavy mass limit the heavy hadrons fall into spin symmetry 
doublets which may be characterized by the spin of the light degrees 
of freedom. Since the heavy quark spin decouples, the total angular 
momentum of the light degrees of freedom becomes a good quantum number. 
Hence the spin symmetry doublets of heavy hadrons are the ones with 
total angular momentum $j + 1/2$ and $j - 1/2$ ($j = 1,2,3,\dots$), 
where $j$ is the angular momentum of the light degrees of freedom.  

For the mesons the ground state spin symmetry doublet are the heavy 
pseudoscalar mesons ($0^-$ states) and the corresponding vector meson 
states ($1^-$ states). For the ground state baryons, the spin of the light 
degress of freedom can either be $j=0$ or $j=1$. For $j=0$ the 
corresponding baryon has spin $1/2$ and is called $\Lambda_Q$, 
and it is the simplest 
object from the point of view of heavy quark symmetry: The spin symmetry 
doublet are the two polarization directions of the $\Lambda_Q$. For 
$j=1$ the baryons can have either spin 
$1/2$ (in which case they are called $\Sigma_Q$) or spin $2/3$ (in 
which case they are called $\Sigma_Q^*$), and hence $\Sigma_Q$ and 
$\Sigma^*_Q$ form another spin symmetry doublet of heavy baryons.   

Spin symmetry has some consequences for  transition 
matrix elements, and we shall consider this here for mesons only. 
It is convenient to represent the mesons by
representation matrices carrying a heavy quark spinor index $A$ and a
light quark index $\alpha$. In fact, the matrix
\begin{equation} \label{mesonrep0}
H(v) = H_{A \alpha} (v) = \frac{1}{2} \sqrt{m_H} \gamma_5
       (\fmslash{v}-1)
\end{equation}
represents the correct coupling of the heavy quark and the light 
degrees of freedom (also carrying spin 1/2) to a pseudoscalar meson 
of total spin 0. The heavy quark is on shell in the limit $m_H \to \infty$,
thus we must require
\begin{equation}
(\fmslash{v}-1) H(v) = 0
\end{equation}

Likewise, the representation for a heavy vector meson is 
\begin{equation} \label{mesonrep1}
H^*(v,\epsilon) = H_{A \alpha}^* (v,\epsilon) = 
\frac{1}{2} \sqrt{m_H} \fmslash{\epsilon}
       (\fmslash{v}-1)
\end{equation}

Rotations of the heavy quark spin rotate the heavy pseuodoscalar
mesons into the corresponding heavy vector mesons and vice versa.   
For a $90^\circ$ rotation of the heavy quark spin around the axis 
$\epsilon$ in a heavy pseudoscalar meson we obtain
\begin{equation} 
W_v (\epsilon, 90^\circ) H(v) = \gamma_5 \fmslash{v} \fmslash{\epsilon} H(v)
= \frac{1}{2} \sqrt{m_H} \fmslash{\epsilon}
  (\fmslash{v}-1) 
\end{equation}
which is the representation matrix of a heavy vector meson.

One may work out the group theory of heavy quark symmetries in more
detail and study their consequences for transition matrix elements 
\cite{HQSymm}. 
Without going into details, the final result is the 
analogue of the Wigner Eckhard theorem. If ${\cal H} (v)$ denotes either
$H(v)$ or $H^*(v,\epsilon)$ and if $| {\cal  H} (v) >$ denotes the
corresponding state in the heavy mass limit, one finds
\begin{equation} \label{wet}
< {\cal H} (v^\prime) | \bar{h}_{v^\prime} \Gamma h_v
                      | {\cal  H} (v) > =
\xi (v \cdot v^\prime ) \mbox{Tr }
\left\{ \overline{{\cal H}} (v^\prime) \Gamma {\cal H} (v) \right\}
\end{equation}
where $\Gamma$ is some arbitrary combination of Dirac matrices and 
$\xi (v\cdot v^\prime)$ is a nonperturbative form factor, the so-called
Isgur Wise function. 

Eq.(\ref{wet}) is the main result of heavy quark symmetry in the
mesonic sector, since it relates {\it every} matrix element of
bilinear heavy to heavy currents between two heavy mesons
to the Isgur Wise function $\xi (v \cdot v')$. Furthermore, 
since the current
\begin{equation}
j_\mu =  \bar{h}_{v} \gamma_\mu h_v
\end{equation}
generates the heavy flavour symmetry, we have a normalization statement
for the Isgur Wise function
\begin{equation} \label{xinorm}
\xi (v \cdot v^\prime = 1 ) = 1
\end{equation}
Note, finally, that the Isgur Wise function in a group theoretical
language is just the reduced matrix element which is universal for
the whole spin flavour symmetry multiplet. The trace in
(\ref{wet}) in the language of the Wigner Eckart theorem 
is the Clebsch Gordan coefficient which is entirely
determined by the current operator and the states of the multiplet.

Furthermore, since the spin symmetry violating terms are the ones proportional 
to the ``strong Bohr magneton'' $g/(2m_Q)$ one would expect that the splitting 
between the partners within a spin symmetry doublet scales as $1/m_Q$. 

For mesons we may consider the quantity 
\begin{eqnarray} \label{mesonsplit}
\Delta &=& M^2 (1^-) - M^2 (0^-) = (M (1^-) + M (0^-))(M (1^-) - M (0^-))
\\ \nonumber        &\approx& 2m_Q (M (1^-) - M (0^-)) 
\end{eqnarray}
which we should expect to be a constant in the heavy quark systems. As can be 
seen in table \ref{masstab1}, $\Delta$ indeed turns out to be constant in the 
$B$ and $D$ meson systems, but to some surprise one obtains also the 
same constant looking into light quark systems. 
\begin{table}
\begin{center}
\begin{tabular}{|c|c|}
\hline
System & $\Delta$ in GeV${}^2$ \\
\hline
$(B^*, B)$    & 0.53 \\
$(D^*, D)$    & 0.54 \\
$(K^*, K)$    & 0.55 \\
$(\rho, \pi)$ & 0.57 \\
\hline
\end{tabular}
\end{center}
\caption{Value of the splitting $\Delta$ for the different systems, 
         the data is from the Particle Data Group \protect{\cite{PDG96}}}
\label{masstab1}
\end{table} 

Similarly, for $\Sigma_Q$--like baryons we consider
\begin{equation} \label{barysplit}
\Delta' = M^2 (3/2) - M^2 (1/2) = 
       \approx 2m_Q (M (3/2) - M (1/2)) 
\end{equation}
which again should turn out to be a constant. In table \ref{masstab2}
we list the corresponding quantities. 

\begin{table}
\begin{center}
\begin{tabular}{|c|c|}
\hline
System & $\Delta'$ in GeV${}^2$ \\
\hline
$(\Sigma_b^*, \Sigma_b)$    & 0.65 \\
$(\Sigma_c^*, \Sigma_c)$    & 0.34 \\
$(\Sigma^*, \Sigma)$        & 0.50 \\
$(\Delta, N)$               & 0.64 \\
\hline
\end{tabular}
\end{center}
\caption{Value of the splitting $\Delta'$ for the different systems}
\label{masstab2}
\end{table} 

It is interesting to note that there seems to be a problem with the 
splitting for the heavy $\Sigma_Q$--type baryons, since the $1/m_Q$ 
scaling between the bottom and the charm systems does not seem to be 
satisfied. The data on $\Sigma_b$ is from DELPHI \cite{DELPHIsigmab}
and is up to now only available from conference talks. One would expect 
that the uncertainties in this number are still large, at least much 
larger than the one on the data on the $\Sigma_c$ and $\Sigma_c^*$, 
which is dominated by CLEO data \cite{CLEOsigmac}. However, if one 
takes the numbers 
in the lighter systems serious, hoping for a similar accident as 
for the mesons, the data in the charm system seem to be on the low 
side.  

\section{Applications of Spin Symmetry}
\subsection{Relation between exclusive semileptonic $b \to c$ transitions}
It is well known that heavy quark symmetry allows to relate the 
decays $B \to D \ell \bar{\nu}_\ell$ and $B \to D^* \ell \bar{\nu}_\ell$. 
The relevant matrix elements are the ones involving the left handed
current for a $b\to c$ transition, which are parametrized in general 
by six form factors 
\begin{eqnarray}
&&\langle D (v') | \bar{c} \gamma_\mu b | B(v) \rangle  =  \sqrt{m_B m_D}
 \left[ \xi_+ (y) (v_\mu + v'_\mu)
     + \xi_- (y) (v_\mu - v'_\mu) \right] \\ 
&& \langle D^* (v',\epsilon) | \bar{c} \gamma_\mu b | B(v) \rangle  = 
       i \sqrt{m_B m_{D^*}} 
\xi_V (y) \varepsilon_{\mu \alpha \beta \rho} \epsilon^{*\alpha}
                       v^{\prime \beta} v^\rho  \\ 
&& \langle D^* (v',\epsilon) | \bar{c} \gamma_\mu \gamma_5 b | B(v) \rangle 
   = \sqrt{m_B m_{D^*}}
       \left[ \xi_{A1} (y) (vv'+1) \epsilon^*_\mu
      -  \xi_{A2} (y) (\epsilon^* v)  v_\mu \right. \nonumber \\
   && \qquad \qquad \left.   -  \xi_{A2} (y) (\epsilon^* v)  v'_\mu \right] ,
\end{eqnarray} 
where we have defined $y = vv'$. 

Of particular interest from the point of view of heavy quark symmetry
is the edge of phase space where the final state
$D$ meson is at rest, i.e. the point $y = 1$. At this kinematical 
point the decay rates take the following form
\begin{equation}
\lim_{y \to 1} \frac{1}{\sqrt{y^2 - 1}} \frac{d\Gamma}{dy}
(B \to D^* \ell \bar{\nu}_\ell) = \frac{G_F^2}{4 \pi^3} (m_B - m_{D^*})^2 
m_{D^*}^3 |V_{cb}|^2 |\xi_{A1} (1)|^2 
\end{equation}
and 
\begin{eqnarray}
\lim_{y \to 1} && \left(\frac{1}{\sqrt{y^2 - 1}}\right) \frac{d\Gamma}{dy}
(B \to D \ell \bar{\nu}_\ell) 
\\ \nonumber && = \frac{G_F^2}{48 \pi^3} (m_B + m_D)^2 
m_D^3 |V_{cb}|^2 
\left| \xi_+ (1) - \frac{m_B - m_D}{m_B + m_D} \xi_- (1) \right|^2 
\end{eqnarray}
Assuming that the $b$ and the $c$ quark 
are heavy one may relate all these form factors to a single one, the 
Isgur Wise function as introduced in (\ref{wet})
\begin{equation}
\xi_i (y) = \xi (y) \mbox{ for } i = +,V,A1,A3, \qquad
\xi_i (y) = 0       \mbox{ for } i = -,A2 .
\end{equation} 
Furthermore, again due to heavy quark symmetries, the Isgur Wise function 
is normalized at $y=1$ as $\xi (y=1) = 1$.  

Corrections to this symmetry limit may be systematically accessed using 
HQET. One important result concerning the $1/m_Q$ corrections is  Luke's
theorem \cite{Lu90}, which states that neither the normalization 
of $\xi_{A1}$ nor the one of $\xi_+$ receive corrections linear in 
$1/m_Q$. Hence the leading corrections to these two form factors are 
of the order $1/m_c^2$. Furthermore, radiative corrections have been 
calculated up to next-to-leading order, and one obtains for $\xi_{A1} (1)$
\begin{equation}
\xi_{A1} (1) = \eta_A (1+\delta_{1/m^2}) 
\end{equation}
where $\eta_A$ incorporates the (QCD and QED) radiative corrections
to the axial-vector $b \to c$ current and $\delta_{1/m^2}$ parametrizes 
the corrections of order $1/m_Q^2$. Inserting numbers one finds  
\begin{equation} \label{a1norm}
\xi_{A1} (1) = 0.92 \pm 0.03
\end{equation}
where the uncertainty is entirely due to the parametrization of the 
corrections of order $1/m_Q^2$. 

Similarly, for the decay $B \to D \ell \bar{\nu}_\ell$ one may write 
for the relevant combination of form factors
\begin{equation}
\left| \xi_+ (1) - \frac{m_B - m_D}{m_B + m_D} \xi_- (1) \right| = 
\eta_V (1+ \Delta_{1/m_Q})
\end{equation}
where $\eta_V$ incorporates the (QCD and QED) radiative corrections
to the vector $b \to c$ current and 
$\Delta_{1/m_Q}$ are the $1/m_Q$ corrections induced by $\xi_- (1)$,
which is not protected by Luke's theorem. 
These corrections have been estimated recently \cite{CapNeu} 
\begin{equation} \label{bdest}
\left| \xi_+ (1) - \frac{m_B - m_D}{m_B + m_D} \xi_- (1) \right| = 
0.98 \pm 0.07
\end{equation}

The absolute normalizations (\ref{a1norm}) and (\ref{bdest}) have been 
used to determine the CKM matrix element $V_{cb}$ using the measured 
spectra of both $B \to D^* \ell \bar{\nu}_\ell$ and 
$B \to D \ell \bar{\nu}_\ell$ \cite{neubertVcb}. 
On the other hand, one may use the same
data \cite{CLEODstar,CLEOD} to test the helicity 
structure of the weak $b \to c$ transition 
current, since at $y=1$  $B \to D^* \ell \bar{\nu}_\ell$ is sensitive
to the axial current only and $B \to D \ell \bar{\nu}_\ell$ is sensitive 
to the vector current only. If we modify the weak $b \to c$ current 
by including coupling constants $g_V$ and $g_A$ according to 
\begin{equation} 
\bar{c} \gamma_\mu (1-\gamma_5) b \longrightarrow 
\bar{c} \gamma_\mu (g_V- g_A \gamma_5) b
\end{equation}
the present data allow to constrain the possible values of the ratio
of the coupling constants 
\begin{equation}
\left| \frac{g_A}{g_V} \right| = 1.02 \pm 0.28 
\end{equation}
The large uncertainty in this number is due to the theoretical 
uncertainty in the $1/m_Q$ corrections to $B \to D \ell \bar{\nu}_\ell$
and to the experimental uncertainties also in 
$B \to D \ell \bar{\nu}_\ell$.

\subsection{Polarization of $b$ hadrons from $Z_0$ decay}
There is a large data sample of $b$ hadrons which originates from 
hadronization of $b$ quarks produced from the decay 
$Z_0 \to \bar{b} b$ at LEPI. The interesting feature of these bottom 
quarks is that their weak couplings are such that they are 
produced with a very high polarization 
\begin{equation}
{\cal P} = \left| \frac{g_A g_V}{g_a^2 + g_V^2} \right| \approx 94\% \, , 
\end{equation} 
and thus the question arises how much of this polarization is 
retained in the polarization of the final state $b$ hadrons. 

Of course here the relevant symmetry is the spin symmetry. If 
one assumes that hadronization is a soft process, then we can 
describe it in the limit $m_b \to \infty$, where the spin of 
the heavy quark decouples. This has already the obvious consequence
that the $\Lambda_b$ baryons from $Z_0$ decay should be polarized 
to a similarly high degree as the $b$ quark itself, and the corrections 
should be effects of order $1/m_Q$. We shall give an estimate of this 
effect below. 

However, only one out of ten $b$ quarks hadronise into a $\Lambda_b$ 
baryon and hence much more data is available on mesons, and we shall 
first analyze the situation for mesons along the lines of Falk and 
Peskin \cite{FalkPeskin}. We start from a 
fully polarized $b$ quark and represent 
the 100\% left handed state as $|\Downarrow \, \rangle$. Fragmentation 
means that the heavy quarks gets dressed with light degrees of freedom 
which have to have spin 1/2. Since there is no preferred spin direction 
for the light degrees of freedom, both $|\uparrow \, \rangle$ and 
$|\downarrow \, \rangle$ should have the same probability amplitude. 

From these quark states we can form the following mesonic states
\begin{eqnarray} \label{states}
&& |\Downarrow \, \rangle |\downarrow \, \rangle = 
| B^* (\lambda = -1) \rangle \\ \nonumber 
&& |\Downarrow \, \rangle |\uparrow \, \rangle = 
\frac{1}{\sqrt{2}}\left[ | B \rangle - | B^* (\lambda = 0) \rangle \right]
\end{eqnarray}
where $\lambda$ is the helicity of the $B^*$ meson. 
Since the $B^*$ decays only electromagnetically, it has a very small 
width compared to the mass difference between the $B^*$ itself and
its spin symmetry partner 
$B$, which has an even smaller width. Hence the two meson states 
involved in (\ref{states}) do not overlapp and hence they become incoherent
before any decay can occur. Thus we may obtain from (\ref{states}) the 
following table of probabilities
\begin{eqnarray} \label{probtab}
& P[B] = \frac{1}{4}& \\ \nonumber
  P[B^* (\lambda = -1)] = \frac{1}{2}\, , \,  
& P[B^* (\lambda =  0)] = {\displaystyle \frac{1}{4}}& \, , \, 
  P[B^* (\lambda =  1)] = 0 
\end{eqnarray}

The $B^*$ is identified by its decay $B^* \to B \gamma$ which occurs after 
the time $1/\Gamma (B^*)$. Since the $B$ mesons are pseudoscalar objects, 
no polarization information can be carried by them. One possibility would 
be the angular distribution of the photon emission, for which one obtains
\begin{eqnarray}
&& \frac{d \Gamma}{d \cos \theta} [B^* (\lambda = \pm 1) \to B \gamma] 
\propto \frac{1}{2} (1+ \cos^2 \theta) \\
&& \frac{d \Gamma}{d \cos \theta} [B^* (\lambda = 0) \to B \gamma] 
\propto \sin^2 \theta 
\end{eqnarray}
where the constant of proportionality is the same in both cases and 
$\theta$ is the angle between the boost direction of the $B^*$ and the 
photon momentum. Using the table of probabilities (\ref{probtab}), we
end up with an isotropic distribution which again does not carry any 
polarization information. 

Thus for the mesons the information on the polarization of the initial 
$b$ quark is entirely transferred into the polarization of the emitted 
photon which is indeed left handed. The photon polarization can, however, 
not be measured with any of the LEP detectors, so not much can be done 
for the mesons.

As mentioned above the situtaion is more promising for baryons, in
particular for the $\Lambda_b$. Here one would na\"i vely expect a 
polarization of the order of 90\%, since the $b$ quark polarization should 
be carried over to the $\Lambda_b$ up to corrections of the order $1/m_Q$. 
This expectation is not supported by data \cite{conftalk}, since much lower
values are found experimentally. Thus one needs to analyze the $1/m_Q$ 
effects quantitatively. 

This has been done by Falk and Peskin \cite{FalkPeskin}, 
who discuss the $\Lambda_b$ 
depolarization through $\Sigma_b$ and $\Sigma_b^*$ intermediate 
states, i.e.\ through the process $Z_0 \to \bar{b} b \to \Sigma_b^{(*)}
\to \Lambda_b$. 

In the case of baryons the hadronization process has to dress the $b$ 
quark with light degrees of freedom of either spin $S=0$ or $S=1$, if we 
restrict ourselves to the ground state baryons. Unlike for the simple 
case of mesons here two parameters enter the analysis. First there is 
the relative probability $A$ to have $S=0$ for the light degrees of 
freedom compared to the $S=1$ case, and the second parameter is the 
relative probability $\omega$ to have the light degrees transversely polarized 
$S_3 = \pm 1$ compared to $S_3 = 0$. In terms of these parameters
one may again set up a table of probabilities for the various helicity 
states of the baryons, assuming again a fully left handed polarized 
$b$ quark in the initial state and incoherence of the various states. 
One finds 
\begin{equation}  
\left[ \quad 
\begin{array}{|c||c|c|c|c|}
\hline
\mbox{state}  
&\lambda=-\frac{3}{2}&\lambda=-\frac{1}{2}
                     &\lambda=\frac{1}{2}&\lambda=-\frac{3}{2} 
\vphantom{\int\limits_t^t} \\ 
\hline \hline 
\Sigma^*_b 
&\frac{1}{2} \omega A&\frac{2}{3}(1-\omega)A& \frac{1}{6} \omega A & 0 
\vphantom{\int\limits_t^t} \\
\hline
\Sigma_b
&    --              &\frac{1}{3}(1-\omega)A& \frac{1}{3} \omega A & -- 
\vphantom{\int\limits_t^t}\\
\hline
\Lambda_b
&    --              &           1          &       0              & --       
\vphantom{\int\limits_t^t} \\ 
\hline \end{array} \quad \right] \frac{1}{1+A}
\end{equation}
where the bracket means that all entries should be multiplied by the overall 
normalization $1/(1+A)$. 

Similarly as for the mesons one now has to analyze the subseqent decays
which are the decays $\Sigma^{(*)}_b \to \Lambda_b \pi$. We shall not 
discuss any of the details here and only quote the final result. For 
``reasonable'' values (actually motivated by the Lund string model) 
of $A=0.45$ and $\omega=0$ one finds a significant depolarization of
the $\Lambda_b$ baryons, namly
\begin{equation}
{\cal P} (\Lambda_b @ LEPI) \approx 68\%
\end{equation}
which is still not enough to explain the low experimental values. 
On the other hand, the analysis of Falk and Peskin has to be taken as 
an estimate depending on the two parameters $A$ and $\omega$, and if the 
experimental values remain as low as they are now, some more theoretical 
work is needed.

\subsection{Polarization in $\Lambda_c$ decays} 
Heavy Quark Symmetries also restrict heavy to light tyransitions. 
While for mesons the number of form factors for e.g. $B \to \pi$ and 
$B \to \rho$ transitions is not reduced, some relations may be found 
for baryons. The $\Lambda_Q$ baryons are the simplest objects from the 
point of view of heavy quark symmetry and indeed spin symmetry
imposes interesting constraints. Consider for example the matrix 
element of a current $\bar{q} \Gamma h_v$ between a heavy 
$\Lambda_Q$ and a light spin-1/2 baryon $B_\ell$, where $q$ is
a light quark. This matrix element is described by only two form
factors \cite{MRRpol} according to 
\begin{equation} \label{h2lbary}
\langle B_\ell (p) | \bar{\ell} \Gamma h_v | \Lambda_Q  (v) \rangle =
\bar{u}_\ell (p) \{ F_1 (v \cdot p ) + \fmslash{v} F_2 (v \cdot p) \} \Gamma
u_{\Lambda_Q} (v) .
\end{equation}
Thus in this particular case spin symmetry dratically reduces the
number of independent Lorentz-invariant amplitudes which 
describe the heavy to light transitions.

This has some interesting implications for exclusive semileptonic 
$\Lambda_c$ decays. For the case of a left handed current 
$\Gamma = \gamma_\mu (1-\gamma_5)$, the semileptonic decay 
$\Lambda_c \to \Lambda \ell \bar{\nu}_\ell$ is in 
general parametrized in terms  of six form factors
\begin{eqnarray}
\langle \Lambda (p) | \bar{q} \gamma_\mu (1-\gamma_5) c | \Lambda_c (v) \rangle &=&
\bar{u} (p) \left[ f_1 \gamma_\mu
               + i f_2 \sigma_{\mu \nu} q^\nu
               +   f_3 q^\mu  \right] u(p')  \nonumber \\
&+& \bar{u} (p) \left[ g_1 \gamma_\mu
               + i g_2 \sigma_{\mu \nu} q^\nu
               +   g_3 q^\mu  \right] \gamma_5 u(p') ,
\end{eqnarray}
where $p' = m_{\Lambda_c} v$ is the momentum of the $\Lambda_c$ whereas
$q = m_{\Lambda_c} v - p$ is the momentum transfer. From this one
defines the ratio $G_A / G_V$ by
\begin{equation}
\frac{G_A}{G_V} = \frac{g_1 (q^2 = 0)}{f_1 (q^2 = 0)} .
\end{equation}

In the heavy $c$ quark limit one may relate the six form factors
$f_i$ and $g_i$ ($i=1,2,3$) to the two form factors $F_j$
($j=1,2$)
\begin{eqnarray}
f_1 &=& - g_1 = F_1 + \frac{m_\Lambda}{m_{\Lambda_c}} F_2 \\
f_2 &=& f_3 = -g_2 = -g_3 = \frac{1}{m_{\Lambda_c}} F_2
\end{eqnarray}
from which one reads off $G_A / G_V = -1$. This ratio is accessible
by measuring in semileptonic decays 
$\Lambda_c \to \lambda \ell \bar{\nu}_\ell$ 
the polarization variable $\alpha$
\begin{equation}
\alpha = \frac{2 G_A G_V}{G_A^2 + G_V^2} 
\end{equation}
which is predicted to be $\alpha = -1$ in the heavy $c$ quark limit. 
The subleading corrections to the heavy $c$ quark limit have been 
estimated and found to be small \cite{ML93}
\begin{equation}
\alpha < -0.95 ,
\end{equation}
and recent measurements yield 
\begin{eqnarray} 
\alpha &=& -0.91 \pm 0.49 \quad \mbox{ARGUS \cite{ARGUSalpha}} \\
\alpha &=& -0.89^{+0.17+0.09}_{-0.11-0.05}  \quad \mbox{CLEO\cite{CLEOlamc}}
\end{eqnarray}
and are in satisfactory agreement with the theoretical predictions. 

Recently the CLEO collaboration also measured the ratio of the form factors
$F_1$ and $F_2$, averaged over phase space. Heavy quark symmetries do not 
fix this form factor ratio, at least not for a heavy to light decay, while 
for a heavy to heavy decay the form factor $F_2$ vanishes in the heavy mass 
limit for the final state quark. CLEO measures \cite{CLEOratio}
\begin{equation} \label{F1F2}
\left\langle \frac{F_2}{F_1} \right\rangle_{\mbox{phase space}} = 
     -0.25 \pm 0.14 \pm 0.08
\end{equation}
which is in good agreement with various model estimates.

\subsection{Quark Helicities in $\Lambda_b \to \Lambda \gamma$}
Another interesting application of relation (\ref{h2lbary}) is the 
rare decay $\Lambda_b \to \Lambda \gamma$ which is a flavour changing 
neutral current process of the type $b \to s \gamma$. The interesting
part of this process is its short distance contribution due to the 
effective Hamiltonian
\begin{equation}
H_{eff} = \frac{4 G_F}{\sqrt{2}} V_{ts} V_{tb}^* C_7 {\cal O}_7
\end{equation}
where $C_7$ is some short distance coefficient and 
\begin{equation}
{\cal O}_7 = \frac{e}{32 \pi^2} m_b \bar{s} \sigma_{\mu \nu} 
             (G_V - G_A \gamma_5) b \, F^{\mu \nu}
\end{equation}
where $F^{\mu \nu}$ is the ususal electromagnetic field strength tensor.

The parameters $C_7$, $G_V$ and $G_A$ may be computed in the 
standard model where one finds $C_7 \approx 0.3$ and
$G_V = 1+ m_s/m_b$ and  $G_A = -1 +m_s/m_b$, i.e. the 
$b$ quark is practically right handed. 

It has been speculated that $b \to s \gamma$ may open a window to 
physics beyond the standard model and the value of $C_7$ has already 
been tested to some extent in the decays $B \to X_s \gamma$ and 
also $B \to K^* \gamma$. However, the helicity stucture of the 
effective Hamiltonian cannot be measured in the mesonic decays, 
they will only be accessible in the decays of $\Lambda_b$ baryons. 
In particular, the decay $\Lambda_b \to \Lambda \gamma$ is a good 
candidate, since the decay of the final state $\Lambda$ is self-analyzing. 

The decay  $\Lambda_b \to \Lambda \gamma$ has been analyzed in detail
in \cite{recks}. Apart from the long distance effects, which have 
been estimated to be small, there is also a problem with the 
application of (\ref{h2lbary}). This relation is expected to work 
best in the region of phase space where the final state light
baryon moves slowly in the rest frame of the decaying $\Lambda_b$. 
Unfortunately, the relevant kinematic region of 
$\Lambda_b \to \Lambda \gamma$ is at the opposite side of phase space, 
since $q^2 = 0$ for the real photon implies for the energy $E$ of the 
light baryon $E \approx m_b / 2$ which becomes large in the infinite 
mass limit. However, it has been argued in \cite{recks} that it is 
very likely that relation (\ref{h2lbary}) still holds for the 
decay $\Lambda_b \to \Lambda \gamma$. 

Assuming an unpolarized $\Lambda_b$ one may measure the polarization 
variable $\alpha '$ 
\begin{equation}
\Gamma = \Gamma_0 
\left[ 1 + \alpha^\prime \vec{n} \cdot \vec{S}_\Lambda \right]
\end{equation}
where $\vec{n}$ is the direction of the outgoing $\Lambda$ in the rest 
frame of the $\Lambda_b$. 

Computing the polarization variable using the CLEO mesurement of the 
form factor ratio (\ref{F1F2}) as input one obtains \cite{recks}
\begin{equation}
\alpha^\prime = 0.387 \frac{2G_V G_A}{G_V^2 + G_A^2}  
\end{equation}
which will allow some test of the helicity structure of the effective
short distance Hamiltonian, once enough data on 
$\Lambda_b \to \Lambda \gamma$ becomes available.

\section{Conclusion}
The fact that in the heavy mass limit the spin of the heavy quark 
decouples has many interesting consequences for processes
involving heavy quarks. In the decays one may analyze the 
helicity structure of the transition operator for which the 
standard model makes definite predictions. Analogous to the 
Michel parameter analysis of the $\mu$ and $\tau$ decays 
one may check the left-handedness of the hadronic currents 
of heavy quarks in a model independent way. However, for 
stringent tests one has to wait for the data coming from 
$B$-factories. 

As far as production of heavy hadrons is concerned, LEPI provided 
the interesting possibility to observe the hadronization of highly
polarized $b$ quarks. For the $b$ quarks hadronizing into mesons 
the polarization information is effectively lost, while some 
of the polarization is retained for hadronization into baryons. 
The amount of depolarization for the $\Lambda_b$ seems to be  
quite high, if the experimental values settle in the region as
given in \cite{conftalk}. Unfortunately, for more data on polarized
$b$ quark fragmentation one prabably 
has to wait for the era of polarized hadron 
colliders.


\begin{thebibliography}{99}
\bibitem{reviews}A subjective selection of recent reviews is: \\
               N.~Isgur and M.~Wise: contribution to {\it Heavy Flavors},
               A.~Buras and M.~Lindner (eds.) 
               (World Scientific, Singapore, 1992);
               M.~Neubert, Phys.\ Rept.\ {\bf 245} (1994) 259; 
               I. Bigi, M. Shifman and N. Uraltsev, TPI-MINN-97-02-T, 
               (1997), hep-ph/9703290   
               T.~Mannel,  
               J. Phys. {\bf G 21} (1995) 1007.
\bibitem{HQSymm}
               N. Isgur and M. Wise,
               Phys. Lett. {\bf B232} (1989) 113 and
               {\bf B237} (1990) 527;
               M.~Voloshin and M.~Shifman,
               Sov. J. Nucl. Phys. {\bf 45} (1987) 292  and
               {\bf 47} (1988) 511. 
\bibitem{HQET} B. Grinstein,
               Nucl. Phys. {\bf B339} (1990) 253; 
               H. Georgi,
               Phys. Lett. {\bf B240} (1990) 447; 
               A. Falk, H. Georgi, B. Grinstein and M. Wise,
               Nucl. Phys. {\bf B343} (1990) 1.    
\bibitem{PDG96} Particle Data Group, Phys.\ Rev.\ {\bf D 50} (1996) No.\ 1.
\bibitem{DELPHIsigmab} see e.g. M. Feindt, talk given at the 
               {\it International Europhysics Conference on High-Energy 
                    Physics}, Jerusalem, Israel, 19-26 Aug 1997.  
\bibitem{CLEOsigmac}G. Brandenburg et al. (CLEO Collaboration) 
                Phys.\ Rev.\ Lett.\ {\bf 78} (1997) 2304. 
\bibitem{Lu90}M. Luke,
               Phys. Lett. {\bf B252} (1990) 447.
\bibitem{CapNeu} I. Caprini and M. Neubert 
                 Phys.\ Lett.\ {\bf B380} (1996) 376.
\bibitem{neubertVcb} The latest update on this subject my be found in 
                 M. Neubert,  talk given at the 
               {\it International Europhysics Conference on High-Energy 
                    Physics}, Jerusalem, Israel, 19-26 Aug 1997.  
\bibitem{CLEODstar}B. Barish  et al., (CLEO Collaboration), 
                  Phys. Rev. {\bf D 51} (1995) 1014.
\bibitem{CLEOD}D. Buskulic et al., (ALEPH collaboration), 
               Phys.\ Lett.\ {\bf B395} (1997) 373; 
               M. Athanas et al., (CLEO collaboration)
               Phys.\ Rev.\ Lett.\ {\bf 79} (1997) 2208.
\bibitem{FalkPeskin} A. Falk, M. Peskin 
               Phys. Rev. {\bf D 49} (1994) 3320.
\bibitem{conftalk}P. Br\"uckmann, these proceedings. 
\bibitem{MRRpol}T. Mannel, W. Roberts and Z. Ryzak, 
                Nucl.\ Phys.\ {\bf B355} (1991) 38; 
                Phys.\ Lett.\ {\bf B255} (1991) 593.
\bibitem{ML93}G. Lin and T. Mannel, 
               Phys.\ Lett.\ {\bf B321} (1994) 417.
\bibitem{ARGUSalpha}H. Albrecht, et al. (ARGUS collaboration) 
                     Phys.\ Lett.\ {\bf B326} (1994) 320.
\bibitem{CLEOlamc}T. Bergfeld, et al. (CLEO collaboration)
                   Phys.\ Lett.\ {\bf B323} (1994) 219.
\bibitem{CLEOratio}G. Crawford et al, (CLEO collaboration), 
                    Cornell preprint CLNS 94/1306, CLEO 94/24 (1994).
\bibitem{recks} T. Mannel and S. Recksiegel, 
                Karlsruhe preprint  TTP-97-06, hep-ph/9701399,  
                J. Phys. {\bf G}, in print;  
                Acta Phys.\ Polon.\ {\bf B28} (1997) 2489.
\end{thebibliography}
\end{document}